\documentclass[twoside]{article}
\usepackage[a4paper]{geometry}
\usepackage[latin1]{inputenc} 
\usepackage[T1]{fontenc} 
\usepackage{RR}

\graphicspath{{figures/}{pictures/}{images/}{./}}
\newcommand{\SC}[1]{\autoref{#1}}

\usepackage{amssymb}
\usepackage{mathptmx}
\usepackage[latin1]{inputenc}
\usepackage{textcomp}
\usepackage[T1]{fontenc}

\newcommand{\emphbf}[1]{\emph{\textbf{#1}}}
\usepackage{rotating}

\usepackage{booktabs}
\usepackage{expdlist}

\usepackage{xspace}
\newcommand{\ie}{i.\,e.\xspace}
\newcommand{\eg}{e.\,g.\xspace}
\newcommand{\etal}{et al.\xspace}
\usepackage{tabu}
\usepackage{multirow}

\usepackage[hang,footnotesize]{subfigure}

\usepackage[final]{hyperref}

\newcommand{\mysubject}{Inria Techreport}

\newcommand{\myauthorsimple}{Isenberg et al.}

\makeatletter
\hypersetup{
  pdftitle={\@title},
  pdfsubject={\mysubject},
  pdfauthor={\myauthorsimple},
  pdfcreator={LaTeX2e},
  pdfkeywords={},
  plainpages=false,       
  hypertexnames=false,    
  bookmarksnumbered=false,
  bookmarksopen=true,     
  bookmarksopenlevel=3,   
  pdfpagemode=UseNone,    
  pdfstartview=Fit,
  pdfborder={0 0 0},      
  breaklinks=true,        
  colorlinks=false,       
  nesting=true}
\makeatother

\ifx\pdfoutput\undefined%
  \renewcommand{\pdfbookmark}[3][]{}
\fi%

\RRNo{8580}
\RRdate{August 2014}
\RRauthor{
\href{http://petra.isenberg.cc/}{Petra Isenberg}\thanks{\href{http://petra.isenberg.cc/}{Petra Isenberg} is with Inria, France. E-mail: petra.isenberg@inria.fr\,.}
\and \href{http://tobias.isenberg.cc/}{Tobias Isenberg}\thanks{\href{http://tobias.isenberg.cc/}{Tobias Isenberg} is with Inria, France. E-mail: tobias.isenberg@inria.fr\,.}
\and \href{http://homepage.univie.ac.at/michael.sedlmair/}{Michael Sedlmair}\thanks{\href{http://homepage.univie.ac.at/michael.sedlmair/}{Michael Sedlmair} is with the University of Vienna, Austria. E-mail: michael.sedlmair@univie.ac.at\,.}
\and \href{http://www.csee.umbc.edu/~jichen/}{Jian Chen}\thanks{\href{http://www.csee.umbc.edu/~jichen/}{Jian Chen} is with the University of Maryland, Baltimore County, USA. E-mail: jichen@umbc.edu\,.}
\and \href{http://cs.univie.ac.at/torsten.moeller/}{Torsten M{\"o}ller}\thanks{\href{http://cs.univie.ac.at/torsten.moeller/}{Torsten M{\"o}ller} is with the University of Vienna, Austria. E-mail: torsten.moeller@univie.ac.at\,.}
}

\authorhead{Isenberg et al.}

\RRtitle{Vers une meilleure compr\'{e}hension de la visualisation \`{a} travers l'analyse de mots-cl\'{e}s}
\RRetitle{Toward a deeper understanding of Visualization through keyword analysis}
\titlehead{Toward a deeper understanding of Visualization through keyword analysis}
\RRresume{
Nous pr\'{e}sentons les r\'{e}sultats d'une analyse exhaustive de mots-cl\'{e}s pour 4366 articles de visualisation soumis \`{a} cinq principales conf\'{e}rences de visualisation. Nous d\'{e}crivons les mots cl\'{e}s principaux, domaines th\'{e}matiques, et les tendances historiques sur 10 ans pour deux jeux de donn\'{e}es: (1) les mots-cl\'{e}s standardis\'{e}s de la taxonomie PCS actuellement utilis\'{e}s pour la soumission d'articles \`{a} IEEE InfoVis, IEEE Vis-SciVis, IEEE VAST, EUROVIS, et IEEE PacificVis depuis 2009 et (2) les mots-cl\'{e}s choisis par les auteurs pour les articles publi\'{e}s dans la s\'{e}rie de conf\'{e}rences de visualisation IEEE (appel\'{e}e IEEE VIS depuis 2004). Notre analyse des sujets de recherche en mati\`{e}re de visualisation peut servir de point de d\'{e}part pour (a) aider \`{a} cr\'{e}er un vocabulaire commun pour am\'{e}liorer la communication entre les diff\'{e}rents sous-groupes du domaine de la visualisation, (b) faciliter la compr\'{e}hension des diff\'{e}rences et points communs entre ces diff\'{e}rents sous-groupes, (c) mieux comprendre les nouvelles tendances de recherche \'{e}mergentes, (d) faciliter l'\'{e}tape cruciale consistant \`{a} trouver les bons experts pour la relecture de soumissions, et (e), \`{a} terme, conduire \`{a} une taxonomie compl\`{e}te de la recherche en visualisation. Un r\'{e}sultat suppl\'{e}mentaire et tangible de notre travail est une application qui permet aux chercheurs en visualisation de parcourir facilement les mots-cl\'{e}s utilis\'{e}s dans plus de 2600 articles de la conference IEEE VIS au cours des 10 derni\`{e}res ann\'{e}es. Cette application vise \`{a} faciliter la s\'{e}lection mieux inform\'{e}e et par cons\'{e}quent plus efficace de mots cl\'{e}s pour les futures publications en visualisation.

}
\RRabstract{
We present the results of a comprehensive analysis of visualization paper keywords supplied for 4366 papers submitted to five main visualization conferences. We describe main keywords, topic areas, and 10-year historic trends from two datasets: (1) the standardized PCS taxonomy keywords in use for paper submissions for IEEE InfoVis, IEEE Vis-SciVis, IEEE VAST, EuroVis, and IEEE PacificVis since 2009 and (2) the author-chosen keywords for papers published in the IEEE Visualization conference series (now called IEEE VIS) since 2004. Our analysis of research topics in visualization can serve as a starting point to (a) help create a common vocabulary to improve communication among different visualization sub-groups, (b) facilitate the process of understanding differences and commonalities of the various research sub-fields in visualization, (c) provide an understanding of emerging new research trends, (d) facilitate the crucial step of finding the right reviewers for research submissions, and (e) it can eventually lead to a comprehensive taxonomy of visualization research. One additional tangible outcome of our work is an application that allows visualization researchers to easily browse the 2600+ keywords used for IEEE VIS papers during the past 10 years, aiming at more informed and, hence, more effective keyword selections for future visualization publications.}
\RRmotcle{analyse de mots-cl\'{e}s, th\`{e}mes de recherche, taxonomie, l'histoire de la visualisation, la th\'{e}orie.}
\RRkeyword{Keyword analysis, research themes, research topics, taxonomy, visualization history, theory.}
\RRprojet{Aviz}  

\RCSaclay 

\begin{document}
\makeRR   
\section{Motivation}
One of the main reasons why the field of visualization is such a fascinating field of research is due to its diversity. We not only refer to the diversity of applications, but the diversity of research methods being employed, the diversity of research contributions being made, as well as the diversity of its roots.

{\bf Diversity of roots:} The term \emphbf{visualization} can be understood very broadly, expressing a long history of its use in common language. Therefore, it is not surprising that concepts of visual thinking have penetrated many areas of science, engineering, and philosophy. The field of modern (computer-based) visualization has been greatly influenced by research methods from the fields of numerics and computer graphics, which have given it its birth in 1990. The impact of human-computer interaction affected the birth of the InfoVis community in 1995 and the influence of applied statistics (such as data mining) and cognition has led to the establishment of VAST in 2006.

{\bf Diversity of research methods:} Given its diverse roots, visualizations remains a highly inter-dis\-ci\-pli\-nary field that borrows and extends research methods from other fields. Methods come from fields as diverse as the broader computer science, mathematics, statistics, machine learning, psychology, cognitive science, semiotics, design, or art.

{\bf Diversity of contributions and applications:} Based on these diverse influences, the results of visualization research can be manifold: from engineering solutions to dealing with large data sources (such as real-time rendering solutions, distributed and parallel computing technologies, novel display devices, and visualization toolkits) to understanding design processes (as in perceptual guidelines for proper visual encodings and interaction or facilitating collaboration between different users through visual tools) to scientific inquiries (such as improved understanding of perceptual and cognitive processes).

While all these diverse influences make the field of visualization research an exciting field to be a part of, they also create enormous challenges. There are different levels of appreciation for all aspects of visualization research, communication challenges between visualization researchers, and the challenge of communicating visualization as a research science to the outside. These issues lead, in particular, to the frequently asked question ``what is visualization?''---among funding agencies or even between colleagues. Given our field's broad nature, we need to ask how we can comprehensively describe and summarize all on-going visualization research. These are not just theoretical and philosophical questions, but the answer to these question has many real-world impacts---from such simple (but career-deciding) questions as finding the right reviewers during peer-review to administrative strategic decisions on conference and journal structures and foci. 

So while ``what is visualization?'' is a fundamental question, it is little discussed within our community. In fact, thus far the approaches have mostly focused on understanding some sub-field of visualization (\eg, \cite{Heer:2012:IDV,Shneiderman:1996:EHT,Thomas:2005:IP}) but the question for the broader community has rarely been tackled beyond general textbook definitions (\eg, \cite{Card:1999:RIV}). Those who have approached the problem, did so in a top-down approach. For example, several taxonomies were suggested by experts based on tasks, techniques, or data models (\eg, \cite{Chi:2000:TVT,Shneiderman:1996:EHT,Tory:2004:HFV}). Another way of splitting visualization into more focused areas has been through specific application foci (\eg, VisSec, BioVis, SoftVis, etc.). 

What is missing in this picture is a bottom-up analysis: What types of visualization research are actually happening as expressed by single research contributions in the visualization conferences and journals. Our paper is one of the first steps in this direction. We analyze author-assigned keywords from the three IEEE VisWeek/VIS conferences of the past ten years as well as author-selected taxonomy entries in the submission system for three IEEE VisWeek/VIS conferences, EuroVis, and PacificVis of the past six years. Based on this analysis, we make the following contributions:

{\bf Mapping visualization research:}
In \SC{sec:results}, through the vehicle of keyword analysis, we build a conceptual map of all visualization work as indexed by individual authors. Our main assumption here is that, while each single keyword might be understood in a slightly different way by different researchers, their co-occurrence with other keywords clarifies their meaning, especially when aggregated over many different usages (\ie, many research papers in a major publication venue). This co-occurrence analysis is the basis for deriving clusters and, therefore, research sub-fields. The use of keywords seen over the past ten years also allows us to understand historical trends and we report on the most prominent declining and rising keywords within all of visualization.

{\bf \href{http://www.keyvis.org/}{keyvis.org}:}
In \SC{sec:webpage} we describe a Web-based search tool that (a) makes the keyword meta-data available to a broad set of people, that (b) helps researchers/users of visualizations to quickly find the right papers that relate to a given topic, and that (c) helps visualization researchers find descriptive keywords for their publications.

{\bf Terminology:}
Visualization research is influenced by a diverse set of application domains. The vocabulary of visualization is influenced by all these application areas. However, the resulting diversity of terms are only understood by a small set of visualization researchers and, therefore, hinders the dissemination of research results and insights across all visualization sub-fields. This is very well articulated in the collection of keywords throughout all of visualization in the past ten years. We are the first that collected and ``cleaned'' this data, making it available to the benefit of our community, allowing its systematic analysis. In \SC{sec:forward} we argue for a unification of different vocabulary.

\section{Related Work}

We are not the first to have made an effort to summarize a large set of visualization papers in order to understand topics or trends. One of the earliest such efforts was a summary and clustering of visualization research papers by Voegele \cite{Voegele:1995:ABV} in 1995 in the form of a two-dimensional clustering of all visualization papers up to this point. Other efforts have focused on specific aspects of visualization research. Sedlmair \etal \cite{Sedlmair:2012:DSM}, for example, did a thorough analysis of all design study papers to summarize practices and pitfalls of design study approaches. Further, Lam \etal \cite{Lam:2012:ESI} studied the practice of evaluations in Information Visualization papers which was then extended to include all visualization papers by Isenberg \etal \cite{Isenberg:2013:SRP}. Others have surveyed, for instance, the literature on interactive visualization \cite{Lam:2008:FIC,Yi:2007:TDU}, on tree visualizations \cite{Schulz:2011:TTV}, on quality metrics in high-dimensional data visualization \cite{Bertini:2011:QMH}, on human-computer collaborative problem-solving \cite{Crouser:2012:AFH}, or on visualization on interactive surfaces \cite{Isenberg:1013:VIS}.

In other disciplines, specific techniques have been used to analyze the scientific literature more broadly: to get a better sense of global research trends, links and patterns within the scientific literature.
Co-word analysis is one approach among others (\eg, co-citation analysis) that has tackled the problem by analyzing the scientific literature according to the co-occurrence of keywords, words in titles, abstracts, or even in the full texts of scientific articles \cite{Callon:1991:CAT,Chuang:2013:TMD,Geisler:2000:MST,He:1999:KDT,Law:1988:PMS,Whittaker:1989:CCS}. Callon \etal \cite{Callon:1986:MDS}, in particular, wrote a seminal book on the topic that provides several methods that others have used and extended upon.  

Co-word analysis has been used in different research areas, \eg, polymer chemistry \cite{Callon:1991:CAT}, acid rain research \cite{Law:1988:PMS}, or education \cite{Ritzhaupt:2010:IDE}. Others further restricted the scope of the literature to specific countries, such as Hu and Liu \etal's \cite{Hu:2013:CAL,Liu:2012:CAD} co-word analysis on library and information science in China. The closest co-word analysis studies to our work are Coulter \etal's \cite{Coulter:1998:SES} work on the software engineering community, Hoonlor \etal's \cite{Hoonlor:2013:TCS} general investigation of the computer science literature, and, most recently, Liu \etal's \cite{Liu:2014:CMT} analysis of the human-computer interaction literature. Liu \etal examined papers of the ACM CHI conference from 1994--2013, identified research themes and their evolution, and classified individual keywords as popular, core, or backbone topics. We employ similar approaches as used in particular in Liu \etal's work. Naturally, however, we differ as our focus is on a different research community, visualization, with different keywords, trends, and patterns and and a different historical evolution.

In the visualization and data analysis literature, the closest work to ours is Chuang \etal's \cite{Chuang:2013:TMD} machine learning tool for topic model diagnostics, and G{\"o}rg \etal's \cite{Gorg:2013:CCA} visual text analysis using Jigsaw, and the CiteVis tool \cite{Stasko:2013:CEC}. These lines of work are not per se co-word analyses. However, their data sources also include visualization research papers. In contrast, we primarily focus on the results of our analysis of themes and trends in the visualization literature rather than on the description of any specific tool or algorithm.

\section{Co-Word Analysis of the Visualization Literature}

For our analysis of the visualization research literature we considered several data sources and analyzed two in detail. 

\subsection{Datasets}
We collected the following datasets:\vspace{-.5em}
\begin{description}[\compact\setlabelphantom{A}\setlabelstyle{\bf\itshape}]
\item[Author-assigned keywords:] Keywords freely assigned by the authors to their research paper. We collected this data manually from paper PDFs as the IEEE digital library contained partially incorrect data.
\item[IEEE terms:] These index terms are manually assigned by an IEEE-hired ``university expert'' \cite{IEEE:2014:TFA} using the IEEE taxonomy \cite{IEEE:1998:IT,IEEE:2013:IT}.
\item[INSPEC terms:] These index terms are automatically derived \cite{IEEE:2014:TFA} either using the same IEEE taxonomy (controlled) or without constraints (non-controlled).
\item[PCS taxonomy keywords:] These keywords chosen from a pre-defined visualization keyword taxonomy and are assigned by the authors during the submission of their research paper.
\end{description}

Details of the data collected can be found in \autoref{tab:datasets}. \autoref{tab:top-ten-2004-2013} provides an overview of the top ten keywords from each data source for IEEE VisWeek/VIS papers for 2004--2013. After a first inspection of the IEEE and INSPEC terms, we found these datasets to be unsuitable to answer our research questions and chose not to analyze them further.

\begin{table}[tb]
	\tabulinesep=2pt
  \centering
  \vspace{-\abovecaptionskip}\caption{\label{tab:datasets} Keyword datasets we collected.}\vspace{.7\abovecaptionskip}
  \begin{tabu}[\columnwidth]{@{}%
	  l@{\hspace{5pt}}%
		c@{\hspace{5pt}}%
		c@{\hspace{5pt}}%
		c@{\hspace{5pt}}%
		c@{\hspace{5pt}}|@{\hspace{5pt}}%
		c@{\hspace{5pt}}%
		c@{}}
		\toprule
    conference & author & IEEE & INSPEC & years & PCS & years\\
    \midrule
		IEEE InfoVis & \checkmark & \checkmark & \checkmark & 2004--2013 & \checkmark & 2008--2013\\
		IEEE Vis/SciVis & \checkmark & \checkmark & \checkmark & 2004--2013 & \checkmark & 2008--2013\\
		IEEE VAST & \checkmark & \checkmark & \checkmark & 2006--2013 & \checkmark & 2008--2013\\
		EuroVis &  & & & & \checkmark & 2008--2013\\
		IEEE PacificVis &  & & & & \checkmark & 2009--2013\\
		\bottomrule
    \end{tabu}
\end{table}

\begin{sidewaystable}[p]
  \centering
  \footnotesize
	\tabulinesep=2pt
	\vspace{-\abovecaptionskip}\caption{Top ten keywords \emphbf{only} of IEEE InfoVis, IEEE Vis/SciVis, and IEEE VAST for the different classification schemes; 2004--2013.}\vspace{.7\abovecaptionskip}
  \begin{tabu}{%
	@{}r%
	*1{@{\hspace{2pt}}|@{\hspace{2pt}}X[l]@{\hspace{4pt}}r}%
	*1{@{\hspace{2pt}}|@{\hspace{2pt}}X[1.5,l]@{\hspace{4pt}}r}%
	*1{@{\hspace{2pt}}|@{\hspace{2pt}}X[1.15,l]@{\hspace{4pt}}r}%
	*1{@{\hspace{2pt}}|@{\hspace{2pt}}X[l]@{\hspace{4pt}}r}%
	*1{@{\hspace{2pt}}|@{\hspace{2pt}}X[l]@{\hspace{4pt}}r}%
	@{}}
	\toprule
  \# & \multicolumn{2}{@{}c@{\hspace{2pt}}|@{\hspace{2pt}}}{author keywords} & \multicolumn{2}{@{}c@{\hspace{2pt}}|@{\hspace{2pt}}}{PCS taxonomy (lower level; only 2008--13)} & \multicolumn{2}{@{}c@{\hspace{2pt}}|@{\hspace{2pt}}}{IEEE terms} & \multicolumn{2}{@{}c@{\hspace{2pt}}|@{\hspace{2pt}}}{INSPEC controlled} & \multicolumn{2}{@{}c@{}}{INSPEC non-controlled} \\
	\midrule
  1 & information visualization & 95 & visual knowledge discovery & 296 & data visualization & 685 & data visualisation & 540 & data visualization & 82 \\
  2 & visual analytics & 86 & graph/network data & 273 & visualization & 272 & data visualization & 393 & information visualization & 62 \\
  3 & visualization & 81 & volume rendering & 237 & rendering (comp. graph.) & 176 & rendering (comp. graph.) & 227 & interactive visualization & 41 \\
  4 & volume rendering & 72 & coordinated and multiple views & 233 & visual analytics & 127 & data analysis & 135 & visual analysis & 32 \\
  5 & flow visualization & 48 & user interfaces & 221 & computer graphics & 117 & interactive systems & 118 & data analysis & 26 \\
  6 & interaction & 38 & biomedical and medical visualization & 218 & data analysis & 107 & computational geometry & 111 & volume rendering & 26 \\
  7 & evaluation & 26 & time-varying data & 204 & computational modeling & 104 & graph theory & 67 & GPU & 25 \\
  8 & isosurfaces & 25 & visualization system and toolkit design & 194 & data mining & 103 & pattern clustering & 65 & visual representation & 22 \\
  9 & volume visualization & 24 & high-dimensional data & 191 & displays & 95 & medical image processing & 63 & direct volume rendering & 20 \\
  10 & parallel coordinates & 23 & interaction design & 185 & humans & 95 & feature extraction & 59 & visual exploration & 20 \\
	\bottomrule
  \end{tabu}
  \label{tab:top-ten-2004-2013}
\end{sidewaystable}

\subsection{Keyword Analysis}
Our general approach was first to extensively clean the data and then to analyze similarities and differences of both keyword sets. We largely followed Liu \etal's \cite{Liu:2014:CMT} process for analyzing keyword data for the ACM Conference on Human Factors in Computing Systems (CHI), extending it with expert coding and an analysis of the taxonomy use. 

\subsubsection{Author-Assigned and Expert-Coded Keyword Datasets}
Our collection of papers from the three IEEE VisWeek/VIS conferences (see \autoref{tab:datasets}) contained 1097 published papers (excluding posters). Out of these, 58 contained no author-assigned keywords, yielding a set of 1039 papers we considered in our analysis. 
These papers contained a total of 2823 unique keywords. Next, we engaged in an extensive manual cleaning pass in which we consolidated keywords that were the same but presented either as singulars/plurals, with spelling mistakes, or as acronyms. This yielded a cleaned dataset that contained 2629 unique keywords that occurred a total of 4780 times across all papers.

Next, we engaged in a manual expert coding of these keywords in order to find higher-level clusters of keyword topics in this dataset. All authors of this paper participated in a multi-pass coding of the cleaned keywords and assigned one or more higher-level cluster codes from a freely evolving, personal code set. Next, we calculated the clusters that emerged from each expert's coding, encoded them in individual keyword-coocurrence matrices, and performed a hierarchical clustering (\eg, \cite{Aggarwal:2014:DCA}) on the joint matrix using average linkage with the Bray-Curtis distance metric \cite{Bray:1957:OUF}. This clustering yielded an ordering of keywords that most strongly matched the consensus of each expert's clustering but did not yet yield satisfactory final clusters. We, thus, engaged in a final manual clustering pass in which we labeled the clusters that emerged or broke them apart if the clustering had not provided meaningful distinctions. The resulting set of keywords contained 156 unique higher-level keywords that occurred a total of 4026 times across all papers (this number is lower than for the cleaned data as potentially generated duplicates per paper were removed). Throughout our paper we refer to this keyword set as ``expert keywords.''

\subsubsection{PCS Taxonomy Keyword Dataset}
The PCS taxonomy keyword data consists of the authors' classification of their papers according to a visualization taxonomy (called ``PCS taxonomy'' from now on). This PCS taxonomy was created in its present form in 2009\footnote{The first version of the taxonomy was created by the 2007 EuroVis paper chairs, Ken Museth, Torsten M\"{o}ller, and Anders Ynnerman. This version was iterated by several people over the next two years. After a broad consultation within in the visualization community, the final version was assembled by Hans-Christian Hege, Torsten M\"{o}ller, and Tamara Munzner. This effort was supported by the VGTC---the IEEE Visualization and Graphics Technical Committee.} and was used for the different visualization venues starting that year (for PacificVis starting 2010). For the year 2008 (and 2009 for PacificVis) we mapped keywords from a previous to the current taxonomy. 
We analyzed anonymized data on 3,927 paper submissions that included the submission IDs and keywords, but the titles only of papers that were finally accepted to the conferences. The dataset included 1070 accepted and 2857 rejected papers. We did not have access to author names or any other identifying information for rejected papers. Considering data from both accepted and rejected papers allowed us to analyze the types of topics that the community was working on in a given year. In total, the PCS taxonomy includes 14 higher-level categories and 127 lower-level keywords (4--17 per higher-level category. In addition, the dataset also included the paper submission types (one of \emph{technique paper}, \emph{system paper}, \emph{application\discretionary{/}{}{/}design study paper}, \emph{evaluation paper}, or \emph{theory\discretionary{/}{}{/}model paper}) for IEEE InfoVis, IEEE Vis/SciVis, IEEE VAST, and EuroVis, but not for PacificVis.

\begin{figure}[tb]
  \centering
  \includegraphics[width=0.65\columnwidth]{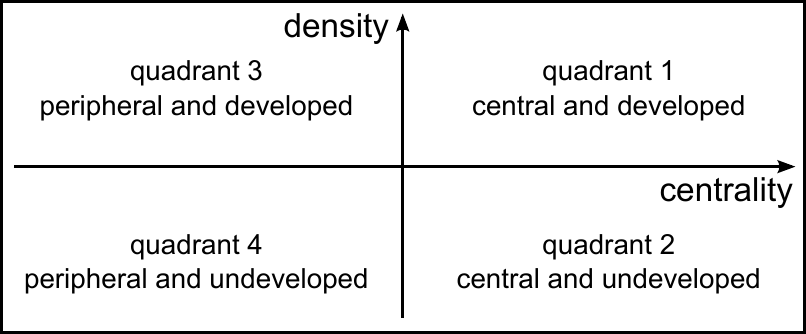}
  \caption{Strategic diagram to characterize the topic clusters (after \cite{He:1999:KDT}).} 
  \label{fig:strategic-diagram}
\end{figure}

\subsubsection{Analysis Process}
To analyze these keyword datasets (cleaned \emphbf{author}-assigned keywords, \emphbf{expert} keywords, \emphbf{PCS} keywords) we first filtered each dataset. We removed keywords that occurred less than a minimum threshold and also excluded higher-level terms as outlined in \autoref{tab:analysismetrics}. Next, we generated document-keyword matrices for each dataset with the keywords as variables (rows) and documents as observations (columns). Each cell contained a 0 if a keyword was not present in a paper and a 1 if it was. On each matrix, we performed a correlation computation using Python's NumPy \texttt{corrcoef} function that yielded a correlation matrix holding keyword-keyword correlation coefficients. On each correlation matrix we performed a hierarchical clustering using Ward's method \cite{Ward:1963:HGO} and a squared Euclidean distance metric. We also generated a keyword network in which two keywords were linked if their correlation was \textgreater\ 0 and each link was assigned its respective correlation value. From this network we computed the density of each cluster with the median of all inter-cluster links and the centrality by computing the sum of square of all links from within a cluster to another cluster. We plotted centrality and density in \emphbf{strategic diagrams} \cite{Callon:1991:CAT,Coulter:1998:SES,He:1999:KDT,Hu:2013:CAL,Liu:2012:CAD,Liu:2014:CMT}. These diagrams distinguish four quadrants (\autoref{fig:strategic-diagram}) that characterize the different clusters based on their centrality within the research domain and on how developed they are. The diagram axes are centered on the median of the observed density and centrality values.

\begin{figure}[tb]
  \centering
  \includegraphics[width=.8\columnwidth]{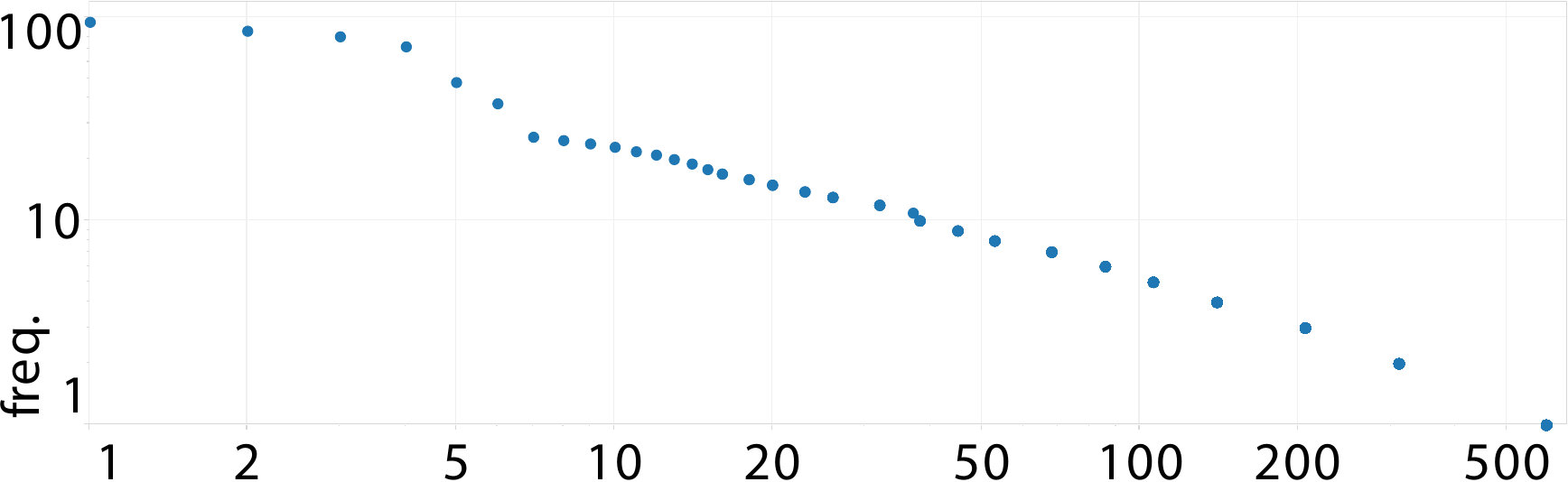}
  \caption{The frequency of the author-assigned keywords (2004--2013, published VIS papers), sorted by their rank (both axes log-transformed).}
  \label{fig:author-frequency-rank}
\end{figure}

We found that the frequency distribution of the author-assigned keywords followed a power law ($\alpha=1.66$, $R^2 =0.80$) as is evident from the linear shape of the log-transformed frequency over rank plot (\autoref{fig:author-frequency-rank}). Similar to the CHI keywords \cite{Liu:2014:CMT}, this indicates that the research structure within visualization is a scale-free network. This means that the research network includes a small number of popular nodes that represent topics on which researchers in the field concentrate. These central nodes serve as hubs to which other research clusters are connected \cite{Liu:2012:CAD,Liu:2014:CMT}. 
In contrast to the author-assigned keywords, the distributions of expert keywords and PCS taxonomy keywords do not follow a power law distribution. For the PCS taxonomy, for example, this means that we do \emphbf{not} observe that the majority of keywords would rarely get used at all, which indicates that the present taxonomy does not contain a majority of topics which would be irrelevant.

\newlength{\lengthofone}
\settowidth{\lengthofone}{1}
\begin{table}[tb]
  \vspace{-\abovecaptionskip}\caption{\label{tab:analysismetrics} The analyzed keyword datasets with their occurrence thresholds applied to keywords, number of remaining keywords analyzed, and target number of clusters set for the cluster analysis.}\vspace{.7\abovecaptionskip} 
	\tabulinesep=2pt
  \centering
  \begin{tabu}[\columnwidth]{@{}%
	  l@{\hspace{3pt}}%
		c@{\hspace{3pt}}%
		l@{}%
		p{57mm}@{\hspace{2pt}}%
		l@{}%
		c@{\hspace{3pt}}%
		c@{}}
		\toprule
    dataset & occur. & \multicolumn{3}{@{}c@{}}{excluded terms} & remaining & target\\
		        & thresh. & & & & keywords & clusters\\ 
    \midrule
		author & \hspace{\lengthofone}6 & \multirow{2}{3ex}{\hspace{-2mm}\bigg\}\bigg\{} & \multirow{2}{\linewidth}{visualization, information visualization, scientific visualization, visual analytics} & \multirow{2}{2.5mm}{\bigg\}} & 101 & 16\\
		expert & 10 & &  & & 101 & 16\\
		PCS    & \hspace{\lengthofone}0  & \multicolumn{3}{@{}c@{}}{none}  & 127 & 15\\
		\bottomrule
  \end{tabu}
\end{table}

\section{Results}
\label{sec:results}
Two main research questions drove our analysis of the data. In particular, we were interested in understanding major research themes and their relationship to other themes (\SC{sec:majorTopics}) and the importance and evolution of individual keywords (\SC{sec:individualkeywords}).

\subsection{Analysis of Major Topic Areas}
\label{sec:majorTopics}
To understand major themes we analyzed the results of the clustering and the generated network graphs. We first report on the results per keyword set and then discuss how these results relate with each other.

\subsubsection{Author-Assigned Keywords}

\begin{table*}[tb]
  \centering
  \footnotesize
	\tabulinesep=2pt
	\vspace{-\abovecaptionskip}\caption{Cluster result for author-specified keywords. Keywords are sorted by frequency with the two most frequent keywords highlighted in bold.}\vspace{.7\abovecaptionskip}
  \begin{tabu}{%
	@{}c%
	*1{@{\hspace{2pt}}|@{\hspace{2pt}}X[l]}%
	*5{@{\hspace{2pt}}|@{\hspace{2pt}}r}%
	@{}}
	\toprule
	ID & keywords (InfoVis, Vis/SciVis, VAST; 2004--2013) & N & \~{\#} & cw-\# & centr. & dens. \\
	\midrule
  A1 & \textbf{isosurfaces}, \textbf{direct volume rendering}, transfer function, gpu, time-varying data, molecular visualiz., raycasting, sampling & 8 & 11.5 & 1.2 & 0.286 & 0.085 \\
  A2 & \textbf{volume visualization}, \textbf{illustrative visualization}, focus+context techniques & 3 & 16.0 & 3.3 & 0.182 & 0.219 \\
  A3 & \textbf{volume rendering}, \textbf{graphics hardware}, interpolation, visualization systems, level-of-detail, vector field visualization, unstructured grids, astronomy, programmable graphics hardware & 9 & 7.0 & 0.6 & 0.318 & 0.047 \\
  A4 & \textbf{focus+context visualization}, \textbf{coordinated \& multiple views}, interactive visualization, interactive visual analysis & 4 & 7.0 & 0.7 & 0.185 & 0.094 \\
  A5 & \textbf{interaction}, \textbf{sensemaking}, multiple views, network visualization, time series data, social networks, exploratory visualization & 7 & 9.0 & 0.8 & 0.316 & 0.063 \\
  A6 & \textbf{geovisualization}, \textbf{spatio-temporal data} & 2 & 9.5 & 4.0 & 0.165 & 0.449 \\
  A7 & \textbf{graph visualization}, \textbf{clustering}, treemaps, node-link diagrams, hierarchies & 5 & 15.0 & 1.8 & 0.278 & 0.143 \\
  A8 & \textbf{parallel coord.}, \textbf{multi-variate data}, user interfaces, high-dimensional data, scatterplots, visual data mining, user interaction & 7 & 12.0 & 1.3 & 0.218 & 0.091 \\
  A9 & \textbf{evaluation}, \textbf{design}, framework, animation, knowledge discovery, taxonomy, experiment & 7 & 8.0 & 1.0 & 0.196 & 0.081 \\
  A10 & \textbf{focus+context}, \textbf{multi-variate visualization}, uncertainty, classification, dimensionality reduction, time-varying, visual analysis, principal component analysis & 8 & 9.0 & 0.5 & 0.239 & 0.047 \\
  A11 & \textbf{user study}, \textbf{human-computer interaction}, visual knowledge discovery, design study, medical visualiz., collaboration, diffusion tensor imaging, linked views, graph layout, text visualiz., non-photorealistic rendering, tiled displays, social data analysis, bioinformatics, applications of visualiz., glyphs, illustrative rendering, intelligence analysis, geographic visualiz. & 19 & 7.0 & 0.1 & 0.283 & 0.011 \\
  A12 & \textbf{feature extraction}, \textbf{unsteady flow visualization} & 2 & 10.0 & 3.0 & 0.143 & 0.308 \\
  A13 & \textbf{uncertainty visualiz.}, \textbf{perception}, graph drawing, multi-dimen. visualiz., multi-dimen. scaling, data mining, qualit. eval. & 7 & 9.0 & 0.5 & 0.213 & 0.049 \\
  A14 & \textbf{flow visualization}, \textbf{vector fields}, streamlines, 3d vector field visualization & 4 & 8.0 & 2.3 & 0.264 & 0.155 \\
  A15 & \textbf{isosurface extraction}, \textbf{hardware acceleration}, point-based visualization & 3 & 7.0 & 1.3 & 0.126 & 0.176 \\
  A16 & \textbf{topology}, \textbf{comparative visualization}, morse-smale complex, data exploration, contour tree, feature detection & 6 & 7.5 & 0.5 & 0.148 & 0.066 \\
	\bottomrule
  \end{tabu}
  \label{tab:authors-cluster-data}
\end{table*}

\textsc{Cluster Analysis:} We created 16 clusters from the 101 author-assigned keywords. The number of clusters we chose was based on manual inspection of the content validity of the hierarchical clustering result. 
\autoref{tab:authors-cluster-data} summarizes the clusters (A1--A16) that were created. In the table we report the keywords assigned to each cluster, the size of the cluster, the average amount of times each keyword occurred in the whole corpus, the average amount of times the keywords in each cluster co-occurred (CW-frequency), the centrality of the subnetworks created from each cluster, and the density of the subnetworks. Keywords are sorted by number of occurrences in the dataset and the top two keywords are bolded and are used to refer to the clusters in the text. 

\begin{figure}[tb]
  \centering
  \includegraphics[width=\columnwidth]{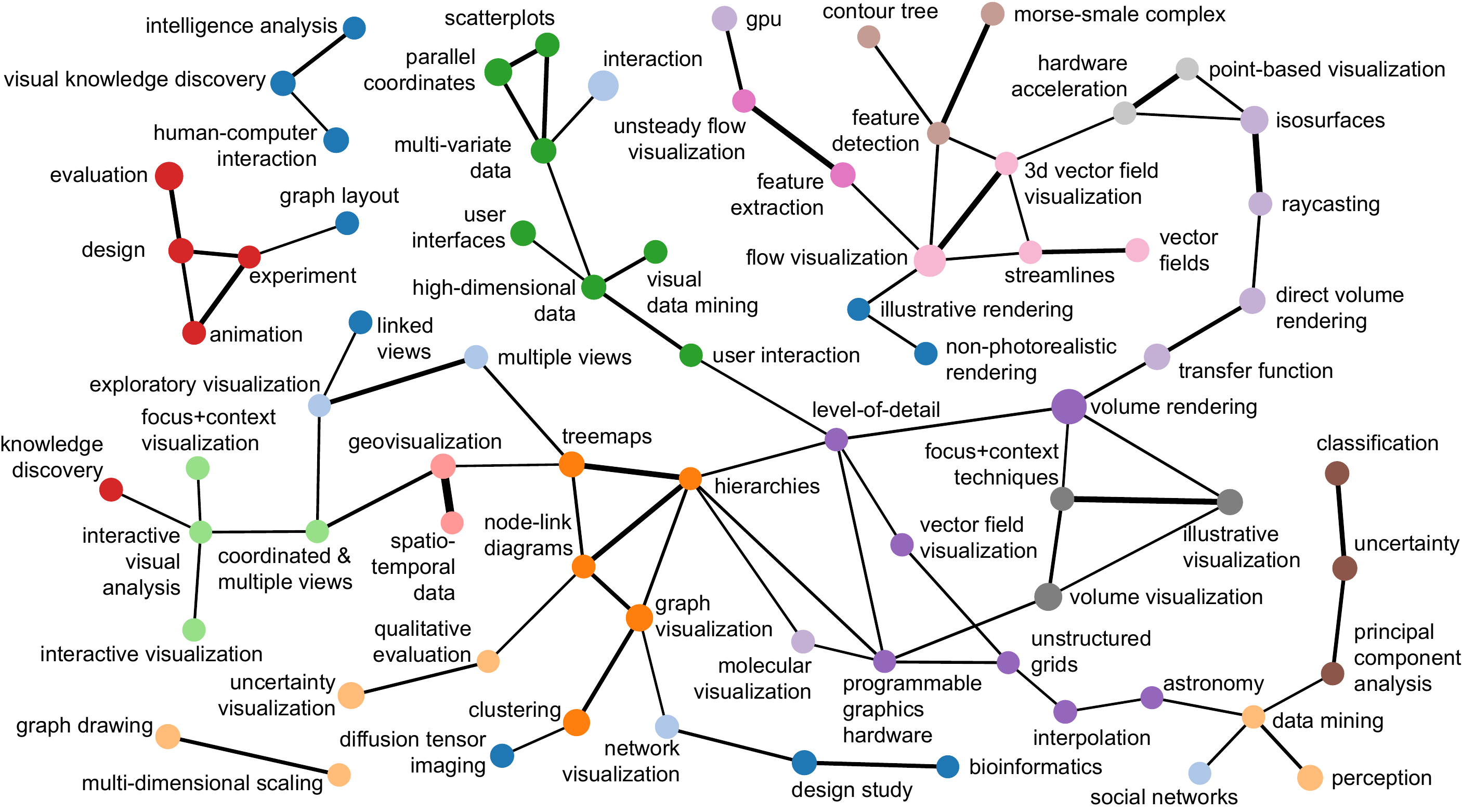}
  \caption{Keyword map from hierarchical clustering of author-assigned keywords; showing only connected nodes with correlation strength $\geq$\,0.13. Circle areas correspond to number of occurrences of the keyword in the dataset, link width corresponds to correlation strength, and color distinguished node clusters. We chose a link threshold based on visual inspection of the resulting graph to generate a manageable and readable layout. Isolated nodes were removed from the image.}
  \label{fig:author-clustergraph}
\end{figure}

By inspecting of the clusters we found that several of them seemed to form more coherent themes than others; many keywords made intuitive sense together, while some other clusters seemed to include outliers. A11, in particular, contains a large number of diverse keywords around applications, encodings, and specific visualization techniques. 
There were only few exceptions where individual keywords seemed slightly out of place. For example, the keywords \emph{astronomy} in A3 or \emph{multi-dimensional scaling} in A13, at first, may seem misplaced. The correlation matrix, however, reveals that \emph{astronomy} was highly positively correlated with \emph{interpolation}, the second most frequent keyword in this cluster, same as \emph{multi-dimensional scaling} with \emph{graph drawing} and \emph{multi-dimensional visualization}. We similarly checked other keywords that seemed out of place and confirmed their placements in the respective clusters to be meaningful with respect to the correlation scores. The network graph in \autoref{fig:author-clustergraph} can also further serve as an analysis aid to check the strongest correlations among keywords.

\autoref{tab:authors-cluster-data} also includes several clusters with similar keywords but using slightly different wording. For example, A1 (\emph{isosurfaces}, \emph{direct volume rendering}), A2 (\emph{volume visualization}, \emph{illustrative visualization}), and A3 (\emph{volume rendering}, \emph{graphics hardware}) center around volumes. Similarly, Clusters A12 and A14 relate to flow visualization.  Both volume and flow visualization have been central topics in the IEEE Vis/SciVis conference, illustrated by the fact that both \emph{volume rendering} and \emph{flow visualization} were the top two non-generic keywords in this dataset. Thus, the emerged clusters around these two general topic areas suggest that the two topics are established and sub-areas of research have emerged. It was more surprising to us that other related terms emerged under different terminology.  For example, three focus+context-related keywords were not clustered together: \emph{focus+context techniques} (A1), \emph{focus+context visualization} (A4), and \emph{focus+context} (A10).  Similarly, \emph{graph visualization} was placed in A7 and \emph{graph drawing} in A13. This suggests that sub-communities within visualization may use slightly different terminology to label similar or related topics.

\begin{figure*}[tb]
  \centering
  \setlength{\subfigcapskip}{-1.5em}%
  \subfigure[\hspace{\textwidth}]{\hspace{.02\textwidth}\label{fig:author-centrality-density}\includegraphics[width=.3\textwidth]{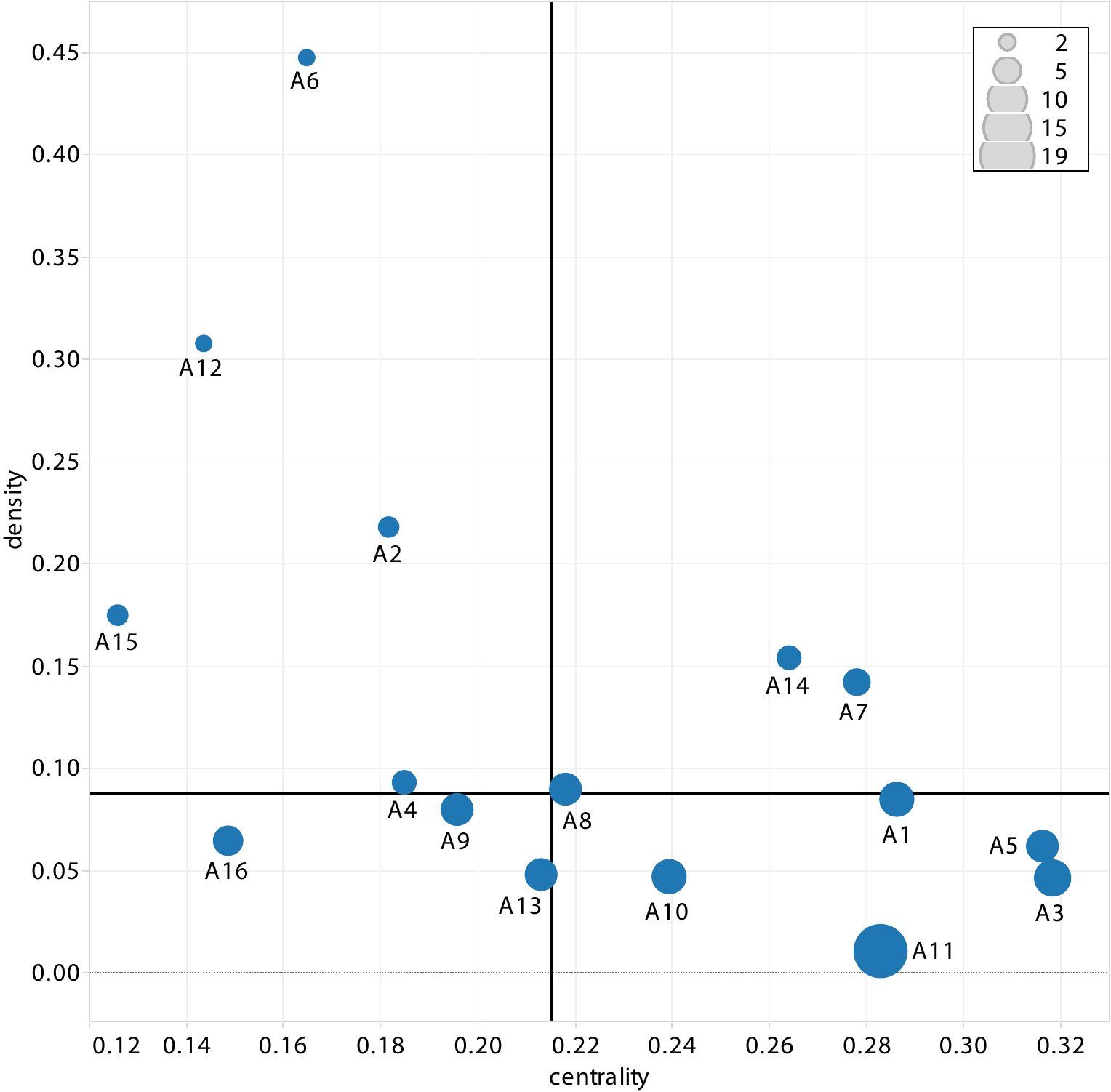}}\hfill%
  \subfigure[\hspace{\textwidth}]{\hspace{.02\textwidth}\label{fig:expert-centrality-density}\includegraphics[width=.3\textwidth]{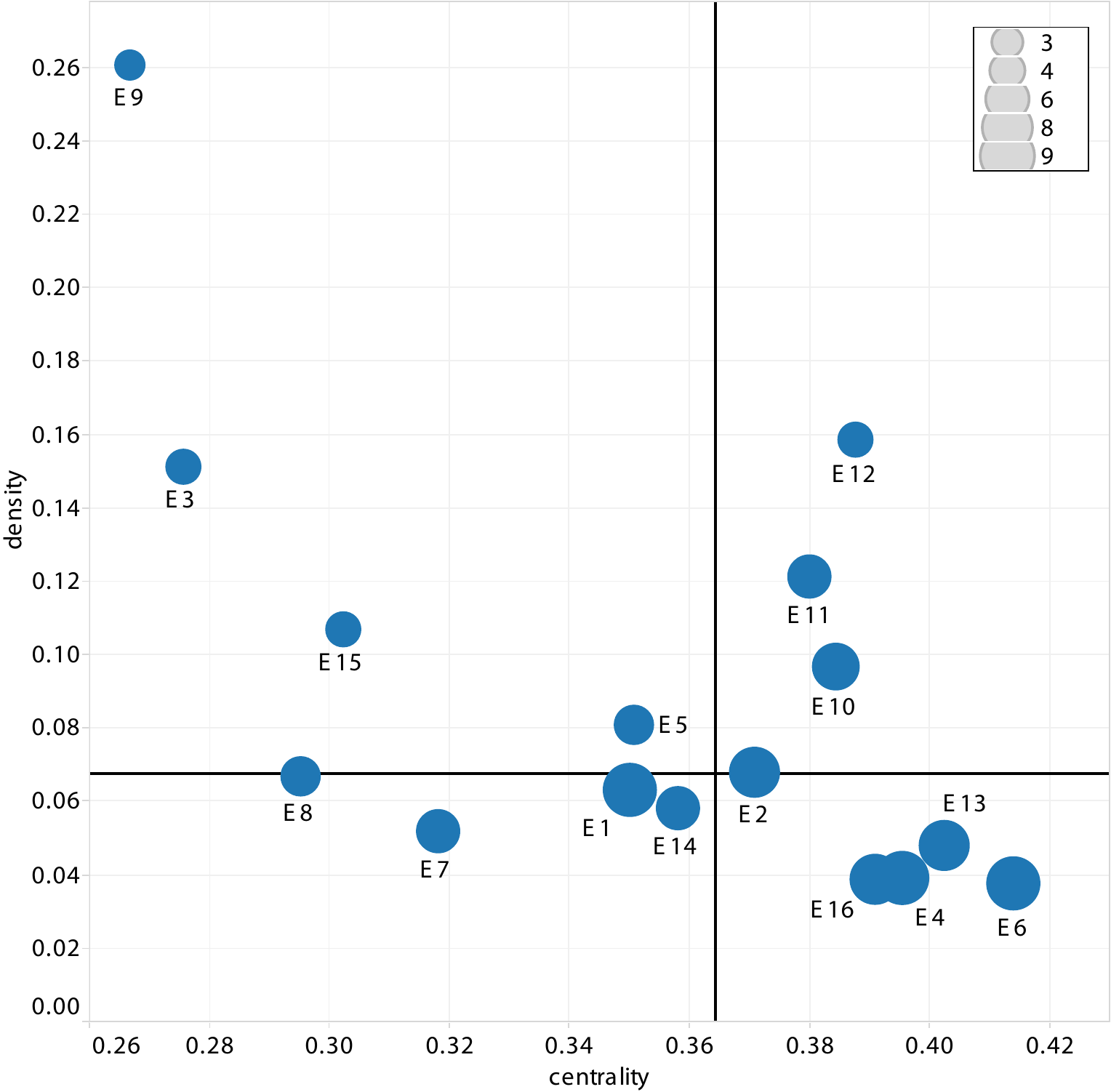}}\hfill%
  \subfigure[\hspace{\textwidth}]{\hspace{.02\textwidth}\label{fig:pcs-centrality-density}\includegraphics[width=.3\textwidth]{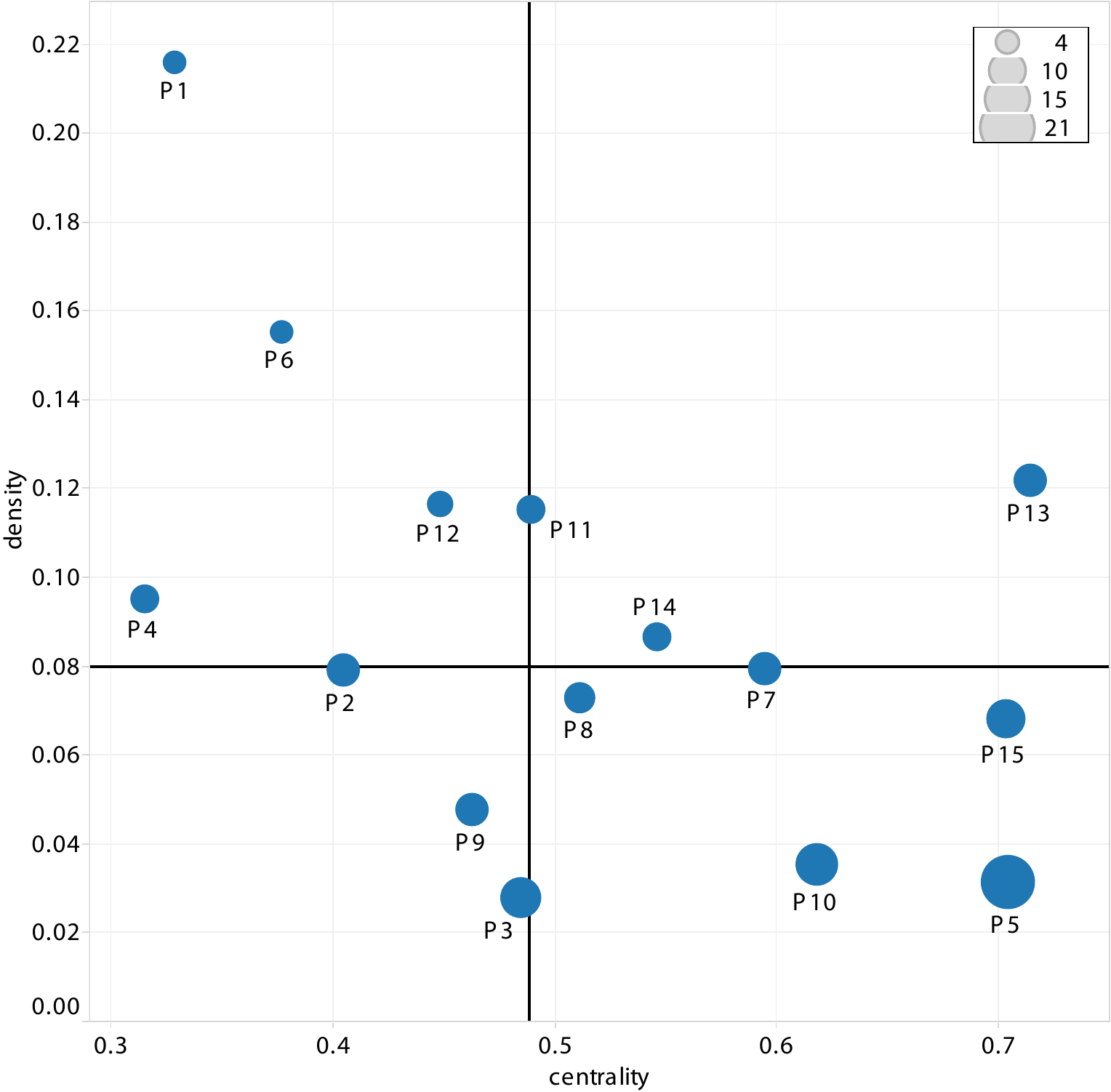}}
  \caption{Strategic diagrams of the cluster results for \subref{fig:author-centrality-density} author-assigned, \subref{fig:expert-centrality-density} expert, and \subref{fig:pcs-centrality-density} PCS keywords. Black lines indicate the medians.}
  \label{fig:strategic-with-data}
\end{figure*}

\textsc{Network Analysis:} Next, we analyzed the strategic diagram in \autoref{fig:author-centrality-density} and the keyword cluster map in \autoref{fig:author-clustergraph}. \autoref{fig:author-centrality-density} shows that topic areas emerge in each of the previously mentioned quadrants. In the following we do not discuss clusters A1, A4, A8, A9, and A13 as their density or centrality measures were very close to the median and, thus, their membership in specific quadrants was not very strong. 

The topics in Quadrant I of the strategic diagram (top right) are considered motor themes or ``mainstream'' topics as they are both internally coherent and central to the research network. In our network, clusters A7 (\emph{graph visualization}, \emph{clustering}) and A14 (\emph{flow visualization}, \emph{vector fields}) are considered motor themes. 

Topics in Quadrant II (bottom right) are A3 (\emph{volume rendering}, \emph{ graphics hardware}), A5 (\emph{interaction}, \emph{sensemaking}), A10 (\emph{focus+context}, \emph{multi-variate visualization}), and A11 (\emph{user study}, \emph{human-computer interaction}). These themes are weakly linked together but present strong ties to other themes (clusters) within the network. They are considered basic and transversal themes.

Quadrant III (top left) includes developed but isolated themes. These are the smaller sub-networks A2 (\emph{volume visualization}, \emph{illustrative visualization}), A6 (\emph{geovisualization}, \emph{spatio-temporal data}), and A12 (\emph{feature extraction}, \emph{unsteady flow visualization}), and A15 (\emph{isosurface extraction}, \emph{hardware acceleration}). Their high density scores indicate that they are relatively strongly tied within themselves---also evident in the network graph in \autoref{fig:author-clustergraph} where these clusters remain mostly tied together through strong (thick) links. While all four subnetworks also had the low centrality scores, they remained connected to the larger central network in \autoref{tab:authors-cluster-data}. 

Finally, Quadrant IV (lower left) contains emerging or declining topics as they are neither very dense nor central to the network.  In particular, A16 (\emph{topology}, \emph{comparative visualization}) lies in this quadrant.

\subsubsection{Expert-coded Keywords}

\textsc{Cluster Analysis:} We created 16 clusters from the 101 expert-coded keywords (after removing 53 whose frequency was \textless\ 10). Again, we chose the cluster number based on a manual inspection of the hierarchical clustering result. \autoref{tab:expert-cluster-data} gives an overview of the created clusters (E1--E16) as well as the same cluster metrics as in \autoref{tab:authors-cluster-data}.

It is important to note that many of the keywords contained in these 101 expert keyword clusters capture topics that were not covered in the cluster analysis of the author keywords. Specifically, 46 of the expert keywords in \autoref{tab:expert-cluster-data} do not contain any of the 101 most frequent author-assigned keywords in \autoref{tab:authors-cluster-data} that we used for author keyword clustering. In other words, these 46 expert keywords capture topics in which none of the contained author keywords occurred more than 5\,\texttimes. Out of these 46, the following ones were among the top two most common themes in one of the expert keyword clusters: \emph{motion} (E3), \emph{data and data management} (E6), \emph{machine learning \& statistics} (E8), \emph{images and image/signal processing} (E8), \emph{surfaces} (E10), \emph{numerical methods/mathematics} (E10), \emph{large-scale data} (E13), \emph{cameras \& views} (E13), \emph{tensors} (E14), \emph{biology} (E15).The top three of these keywords occurred more than 50\,\texttimes\ in our corpus and would not have emerged as important without the expert coding.

\begin{table*}[tb]
  \centering
  \footnotesize
	\tabulinesep=2pt
	\vspace{-\abovecaptionskip}\caption{Cluster result for expert-coded keywords. Keywords are sorted by frequency with the two most frequent keywords highlighted in bold.}\vspace{.7\abovecaptionskip}
  \begin{tabu}{%
	@{}c%
	*1{@{\hspace{2pt}}|@{\hspace{2pt}}X[l]}%
	*5{@{\hspace{2pt}}|@{\hspace{2pt}}r}%
	@{}}
	\toprule
	ID & keywords (InfoVis, Vis/SciVis, VAST; 2004--2013) & N & \~{\#} & cw-\# & centr. & dens. \\
	\midrule
  E1 & \textbf{graphs}, \textbf{visualiz. techniques \& tools}, clustering, hierarchies, layout, charts/diagrams/plots, networks, scatterplots, matrices & 9 & 34.0 & 3.9 & 0.355 & 0.063 \\
  E2 & \textbf{evaluation}, \textbf{perception}, design \& guidelines, visual encoding, color, human factors/hci, art \& aesthetics, provenance \& history & 8 & 28.5 & 3.2 & 0.377 & 0.068 \\
  E3 & \textbf{geography/geospatial/cartography}, \textbf{motion}, spatiotemporal, maps & 4 & 20.5 & 4.0 & 0.276 & 0.152 \\
  E4 & \textbf{displays}, \textbf{toolkits/systems/environments}, interfaces, theory/models/methods, knowledge discovery, cognition, metrics, video \& multimedia, virtual environments & 9 & 29.0 & 1.4 & 0.394 & 0.040 \\
  E5 & \textbf{interaction}, \textbf{analysis process}, multidimensional/multivariate, multiple \& linked views, queries \& search & 5 & 83.0 & 11.5 & 0.350 & 0.081 \\
  E6 & \textbf{data \& data management}, \textbf{dimensionality reduction}, social data/media/vis, databases \& data mining, collaboration, programming, internet/web, scale \& multiscale, curves \& curvature & 9 & 25.0 & 1.5 & 0.413 & 0.038 \\
  E7 & \textbf{applications}, \textbf{text \& documents}, physics, patterns \& pattern detection, semantics/semiotics, storytelling & 6 & 20.5 & 2.2 & 0.317 & 0.052 \\
  E8 & \textbf{machine learning \& statistics}, \textbf{images \& image/signal processing}, uncertainty, segmentation/classification, simulation & 5 & 34.0 & 3.0 & 0.295 & 0.067 \\
  E9 & \textbf{interpolation}, \textbf{sampling}, grids \& lattices & 3 & 14.0 & 4.0 & 0.267 & 0.261 \\
  E10 & \textbf{surfaces}, \textbf{numerical methods/math}, isosurfaces, meshes \& grids, geometric modeling, point based data, raytracing/raycasting & 7 & 32.0 & 4.0 & 0.384 & 0.097 \\
  E11 & \textbf{flow}, \textbf{topology}, vectors, features \& attributes, 3d visualization, streamlines/pathlines/streaklines & 6 & 42.0 & 6.6 & 0.380 & 0.122 \\
  E12 & \textbf{volume rendering}, \textbf{hardware \& computation}, rendering \& illumination, textures & 4 & 68.5 & 14.8 & 0.387 & 0.159 \\
  E13 & \textbf{large scale data}, \textbf{cameras \& views}, resolution/multiresolution, astronomy/astrophysics, computer networks, particles, information theory, level of detail & 8 & 15.0 & 1.1 & 0.403 & 0.048 \\
  E14 & \textbf{medical}, \textbf{tensors}, tractography, glyphs, climate/environment, biomedical & 6 & 20.0 & 1.9 & 0.358 & 0.058 \\
  E15 & \textbf{design studies \& case studies}, \textbf{biology}, molecules, bioinformatics & 4 & 19.0 & 2.2 & 0.302 & 0.107 \\
  E16 & \textbf{time}, \textbf{focus+context}, illustrative visualiz., comparison/comp. visualiz., spatial, events/trends/outliers, animation, comp. graph. & 8 & 22.0 & 1.6 & 0.391 & 0.039 \\
	\bottomrule
  \end{tabu}
  \label{tab:expert-cluster-data}
\end{table*}

\textsc{Network Analysis:} 
The strategic diagram for expert keywords can be seen in \autoref{fig:expert-centrality-density}. From the following discussion we exclude clusters E1, E2, and E8, as their centrality and density scores were very close to median centrality or density and, thus, quadrant membership was not strong for these clusters.

Three central and developed motor themes emerged in Quadrant I (top right): E10 (\emph{surfaces}, \emph{numerical methods/mathematics}), E11 (\emph{flow}, \emph{topology}), E12 (\emph{volume rendering}, \emph{hardware and computation}). All three themes are traditional topics of IEEE Vis/SciVis which also has the longest history of the three IEEE VIS conferences. Hence it is perhaps not surprising that these topics emerged as central and dense. \emph{Flow visualization} and \emph{vector fields} also proved to be a motor theme in the author keywords, further emphasizing E11's place as motor theme.

As undeveloped yet still central themes in Quadrant II (bottom right), clusters E4 (\emph{displays}, \emph{toolkits/systems/environments}), E6 (\emph{data and data management}, \emph{dimensionality reduction}), E13 (\emph{large scale data}, \emph{cameras \& views}), and E16 (\emph{time}, \emph{focus+context}) emerged most clearly. 

Themes in Quadrant III (top left; peripheral to the general research network) were E3 (\emph{geography/geospatial/cartography}, \emph{motion}), E5 (\emph{interaction}, \emph{analysis process}), E9 (\emph{interpolation}, \emph{sampling}), and E15 (\emph{design studies \& case studies}, \emph{biology}). E9 is an extreme case as it exhibits the highest density and lowest centrality score in the network and is also quite small. This theme is thus the most developed but isolated of the themes in this Quadrant. E5, in contrast, was very close to the median for both centrality and density and is thus unspecific.

\begin{figure}[tb]
  \centering
  \includegraphics[width=\columnwidth]{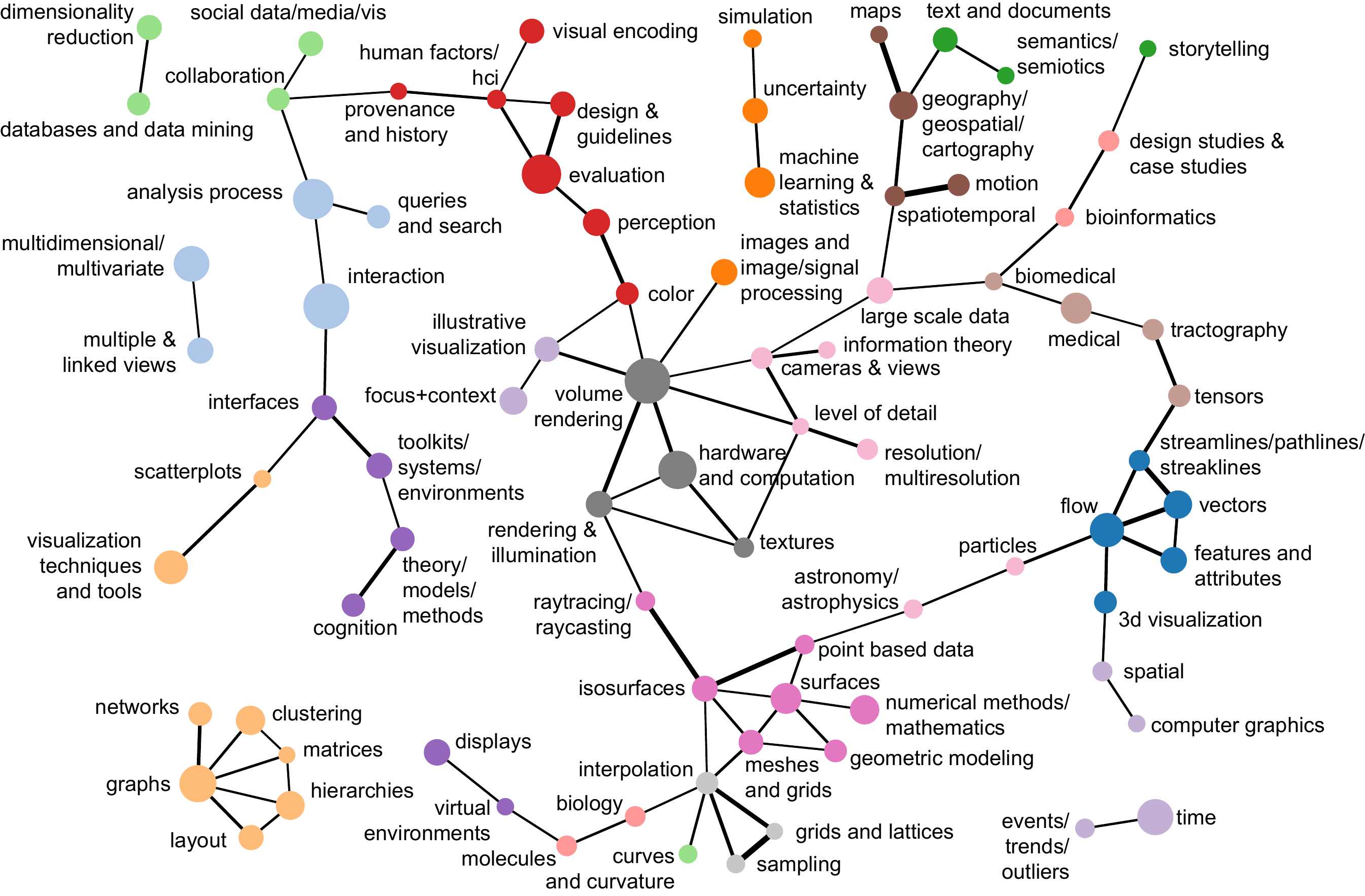}
  \caption{Keyword map from hierarchical clustering of the expert-coded keywords; showing only connected nodes with correlation strength $\geq$\,0.11.}
  \label{fig:expert-clustergraph}
\end{figure}

Among the emerging or declining themes in Quadrant IV (lower left) were clusters E7 (\emph{applications}, \emph{text and documents}) and E14 (\emph{medical}, \emph{tensors}). Interestingly, the graph visualization-related cluster A7 for the author keywords was characterized as a motor theme during the author-keyword analysis but exhibited much lower centrality and lower density for the expert-coded cluster E1. The network diagram in \autoref{fig:expert-clustergraph} gives a visual overview of how words and clusters correlated and how central certain keywords are placed within the general network.

\subsubsection{PCS Taxonomy Keywords}
Before discussing the analysis of PCS keywords, it is important to note that their usage in practice is different than for author keywords. Author keywords are generally used to label a paper and to explain in a few words what research topics it covers---for example, with the intention to make the paper discoverable in searches. While the PCS keywords can be used in the same way, the choices are limited and it is not always possible to find the one keyword that exactly expresses a contribution. 
For the PCS data we analyzed, authors were instructed to choose only a limited subset of PCS keywords (\eg, a primary and up to four secondary keywords). Authors, thus, had to balance their choices among all possible keywords that describe a paper contribution. A fundamental difference to author keywords, however, is that the PCS taxonomy is used not only by the paper authors but also by the reviewers to describe their expertise levels for each topic. The selection of PCS keywords for papers and by reviewers then informs how reviewers are assigned to papers. Since this process is known to many paper authors, PCS keywords are often carefully selected to rank more highly for reviewers with a certain expertise. 

Given these differences in use, PCS keywords may not as accurately reflect the topic of a paper as the author keywords do. Nevertheless, by conducting the same analysis on this keyword set as we did before for author keywords, we can speculate on differences in use as well as on topics that may be missing in each type of keyword set.

Overall, we observed that, similar to author and expert keywords, a small number of PCS keywords appeared very frequently while a larger number of keywords did not. The five least frequently used PCS keywords for the entire period of the dataset were \emph{volume graphics hardware} (8\,\texttimes), \emph{haptics for visualization} (7\,\texttimes), \emph{embodied/enactive cognition} (6\,\texttimes), \emph{sonification} (5\,\texttimes), and \emph{special purpose hardware} (3\,\texttimes). 

\begin{figure}[tb]
  \centering
  \includegraphics[width=\columnwidth]{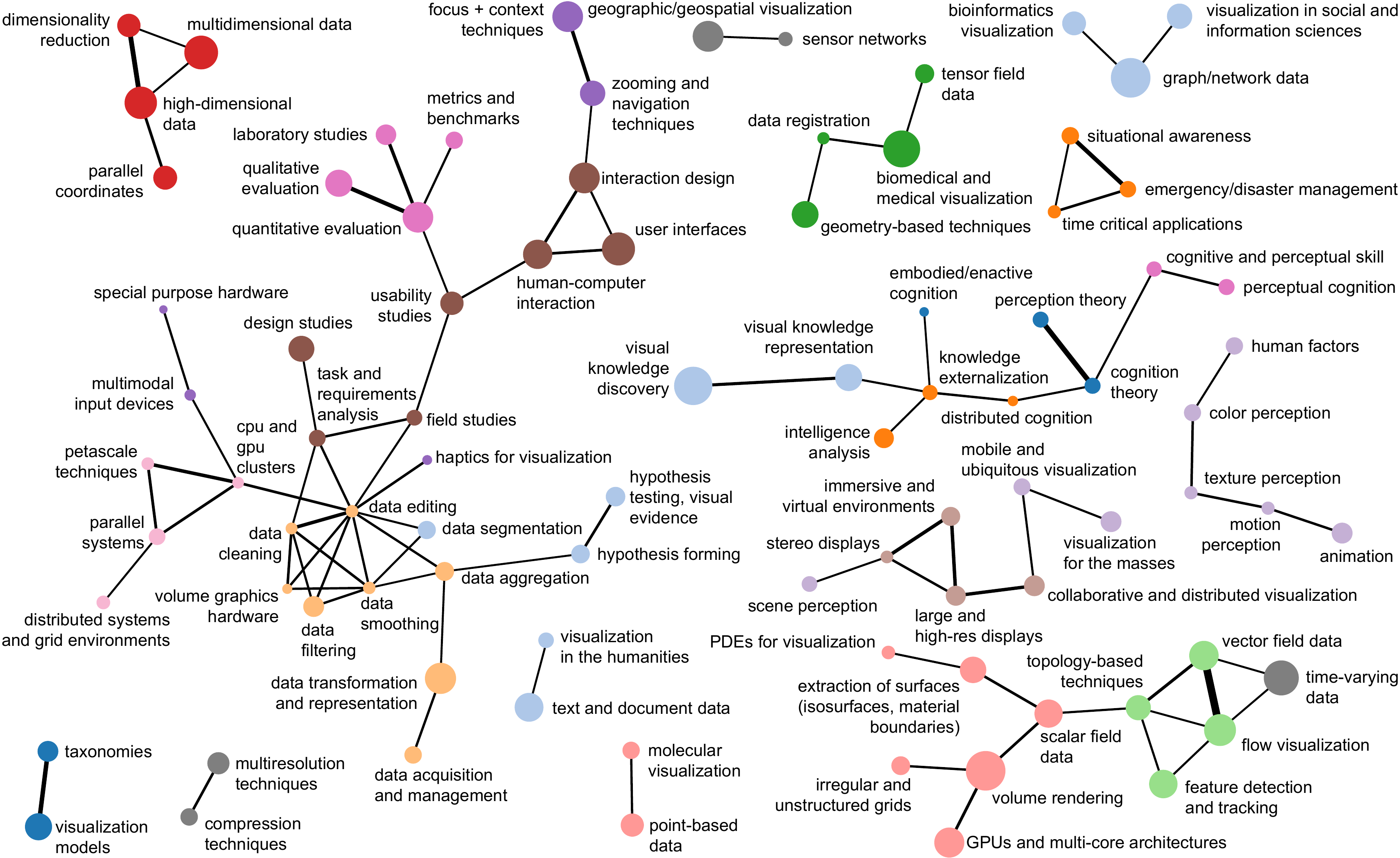}
  \caption{Keyword map from hierarchical clustering for PCS taxonomy keywords; showing only connected nodes with correlation strength $\geq$\,0.135.}
  \label{fig:pcs-clustergraph}
\end{figure}

\textsc{Cluster Analysis:}
The result of the analysis of the PCS taxonomy is reported in the same way as that of the other two keyword sets. \autoref{tab:PCS-cluster-data} has the overview of the generated clusters, their keywords, and metrics. \autoref{fig:pcs-clustergraph} shows a filtered network graph generated from the keyword correlations, while \autoref{fig:pcs-centrality-density} depicts the strategic diagram. 

The cluster analysis includes several interesting observations. First---and perhaps not too surprising---most clusters include keywords that belong to different higher-level taxonomy keywords. For example, \emph{flow visualization} in P1 is considered an application according to the PCS taxonomy, while \emph{vector field data} belongs to the category \emph{spatial data and techniques}. Yet, many clusters include keywords from a predominant subset of the higher level of the PCS taxonomy: P2 (\emph{spatial data and techniques}), P4 (\emph{non-spatial data and techniques}), P6 (\emph{display and interaction technology}), P7 (\emph{visual analytics applications}), P11 (\emph{evaluation}), and P13 (\emph{data handling, processing and analysis}). 

\textsc{Network Analysis:}
Next, we analyzed the strategic diagram in \autoref{fig:pcs-centrality-density}. From the following discussion we exclude P2, P3, P7, and P11, as their density or centrality scores were too close to the media to derive conclusive quadrant memberships.

The most evident motor theme in Quadrant I (top right) was P13 (\emph{data transformation and representation}, \emph{data filtering}). P14 (\emph{focus+context techniques}, \emph{zooming and navigation techniques}) also fell within Quadrant I but was relatively close to median centrality. The network diagram in \autoref{fig:pcs-clustergraph} illustrates the high density and centrality of P13 and its keywords (colored in light orange) well. 

The themes in Quadrant II (bottom right; undeveloped topics that are central to the research network because they are well connected) are P5 (\emph{graph/network data}, \emph{visual knowledge discovery}), P10 (\emph{illustrative visualization}, \emph{animation}), P8 (\emph{scalability issues}, \emph{streaming data}), and P15 (\emph{user interfaces}, \emph{visualization system \& toolkit design}). 

Quadrant III (top left; developed but isolated themes) contained, in particular, P1 (\emph{flow visualization}, \emph{vector field data})---the theme with highest density---and P4 (\emph{multidimensional data}, \emph{high-dimensional data})---the theme with the lowest centrality. The other two themes in Quadrant III are P6 (\emph{collaborative and distributed visualization}, \emph{large and high-res displays}) and P12 (\emph{visualization models}, \emph{taxonomies}. 

Finally, Quadrant IV (bottom left; emerging/declining themes---undeveloped and peripheral to the network) contains P9 (\emph{time-varying data}, \emph{geographic/geospatial visualization}).

\begin{table*}[tb]
  \centering
  \footnotesize
	\tabulinesep=2pt
	\vspace{-\abovecaptionskip}\caption{Cluster result for PCS taxonomy keywords. Keywords are sorted by frequency with the two most frequent keywords highlighted in bold.}\vspace{.7\abovecaptionskip}
  \begin{tabu}{%
	@{}c%
	*1{@{\hspace{2pt}}|@{\hspace{2pt}}X[l]}%
	*5{@{\hspace{2pt}}|@{\hspace{2pt}}r}%
	@{}}
	\toprule
	ID & keywords (EuroVis, PacificVis, InfoVis, Vis/SciVis, VAST; 2008--2013) & N & \~{\#} & cw-\# & centr.\ & dens. \\
	\midrule
  P1 & \textbf{flow visualization}, \textbf{vector field data}, feature detection and tracking, topology-based techniques & 4 & 210.0 & 26.3 & 0.328 & 0.216 \\
  P2 & \textbf{volume rendering}, \textbf{gpus and multi-core architectures}, scalar field data, extraction of surfaces (isosurfaces,  material boundaries), point-based data, irregular and unstructured grids, molecular visualization, pde's for visualization & 8 & 143.0 & 11.5 & 0.404 & 0.079 \\
  P3 & \textbf{biomedical and medical visualization}, \textbf{visualization in physical sciences and engineering}, geometry-based techniques, multi-field,  multi-modal and multi-variate data, uncertainty visualiz., mathematical foundations for visualization, glyph-based techniques, tensor field data, volume modeling, visualiz.\ in mathematics, data registration, sonification & 12 & 124.5 & 4.1 & 0.484 & 0.028 \\
  P4 & \textbf{multidimen.\ data}, \textbf{high-dimen.\ data}, parallel coordinates, dimensionality reduction, statistical graphics, tabular data & 6 & 125.0 & 11.9 & 0.314 & 0.095 \\
  P5 & \textbf{graph/network data}, \textbf{visual knowledge discovery}, coordinated \& multiple views, time series data, text \& document data, data clustering, visual knowledge represent., hierarchy data, visualiz.\ in social \& information sciences, bioinformatics visualiz., visual analysis models, integrating spatial \& non-spatial data visualiz., machine learning, pixel-oriented techniques, software visualiz., hypothesis testing,  visual evidence, hypothesis forming, business \& finance visualiz., data segmentation, data fusion \& integration, visualiz.\ in the humanities & 21 & 119.0 & 8.1 & 0.704 & 0.032 \\
  P6 & \textbf{collaborative \& distributed visualization}, \textbf{large \& high-res displays}, immersive \& virtual environments, stereo displays & 4 & 74.5 & 3.8 & 0.376 & 0.155 \\
  P7 & \textbf{intelligence analysis}, \textbf{situational awareness}, emergency/disaster management, knowledge externalization, network security and intrusion, time critical applications, privacy and security, distributed cognition & 8 & 29.5 & 1.4 & 0.594 & 0.080 \\
  P8 & \textbf{scalability issues}, \textbf{streaming data}, data warehousing,  database visualization and data mining, parallel systems, distributed systems and grid environments, petascale techniques, cpu and gpu clusters & 7 & 47.0 & 2.5 & 0.511 & 0.073 \\
  P9 & \textbf{time-varying data}, \textbf{geographic/geospatial visualization}, visualization in earth,  space,  and environmental sciences, multiresolution techniques, view-dependent visualization, compression techniques, terrain visualization, sensor networks & 8 & 98.5 & 4.8 & 0.462 & 0.048 \\
  P10 & \textbf{illustrative visualization}, \textbf{animation}, visualization for the masses, aesthetics in visualization, multimedia (image/video/music) visualization, human factors, color perception, mobile and ubiquitous visualization, scene perception, visualization in education, motion perception, texture perception, attention and blindness & 13 & 51.0 & 1.7 & 0.617 & 0.035 \\
  P11 & \textbf{quantitative eval.}, \textbf{qualitative eval.}, laboratory studies, metrics \& benchmarks, percept.\ cognition, cognitive \& percept.\ skill & 6 & 69.5 & 10.0 & 0.489 & 0.116 \\
  P12 & \textbf{visualization models}, \textbf{taxonomies}, cognition theory, perception theory, embodied / enactive cognition & 5 & 42.0 & 5.8 & 0.448 & 0.117 \\
  P13 & \textbf{data transformation and representation}, \textbf{data filtering}, data aggregation, data acquisition and management, data smoothing, data editing, data cleaning, volume graphics hardware & 8 & 36.0 & 5.1 & 0.714 & 0.122 \\
  P14 & \textbf{focus+context techniques}, \textbf{zooming and navigation techniques}, manipulation and deformation, multimodal input devices, haptics for visualization, special purpose hardware & 6 & 44.0 & 3.6 & 0.546 & 0.087 \\
  P15 & \textbf{user interfaces}, \textbf{visualiz.\ system \& toolkit design}, interact.\ design, human-computer interact., visual design, design studies, usability studies, design methodologies, task and requirements analysis, presentation/production/dissemination, field studies & 11 & 159.0 & 11.1 & 0.703 & 0.069 \\
	\bottomrule
  \end{tabu}
  \label{tab:PCS-cluster-data}
\end{table*}

\subsubsection{Relationship between Expert and PCS Keywords}

As both expert and PCS keywords present a higher-level assessment of the visualization topics, we were interested in analyzing their relationship.  Yet, we found it difficult to generate an exact match from PCS to expert keywords and vice versa---in part because PCS and expert keywords were not always on the same levels. For example, many specific application domains in the expert keywords could not be matched well because we had matched many applications into a generic \emph{applications} group due to their relatively infrequent use. On the other hand, in the expert coding we had explicitly labeled specific application domains that only existed as relatively large groups in the PCS taxonomy (\eg, expert keywords \emph{physics}, \emph{material science}, \emph{automation}, \emph{traffic} vs.\ \emph{visualization in physical sciences and engineering}; or expert keywords \emph{astronomy/astrophysics}, \emph{climate/environment}, \emph{earth sciences} vs.\ \emph{visualization in earth, space, and environmental sciences}).

We also found the higher-level PCS category \emph{data handling, processing and analysis} to contain very fine-grained keywords which, in our expert coding, were simply mapped to \emph{data and data management}. Similarly, we had merged all author keywords related to \emph{scalability issues} and \emph{petascale techniques} into one expert-keyword: \emph{large-scale data}. On the other hand, we also found PCS categories that relate to several expert keywords and are, thus, rather broad. For example the keyword \emph{visual design} relates to the expert keywords \emph{layout}, \emph{visual encoding},  and \emph{design \& guidelines}. 

Despite these differences in categorization, we found a number of expert keywords that we could not match well in the PCS taxonomy but did appear in our dataset 20\,\texttimes\ or more. These keywords are: \emph{rendering \& illumination}, \emph{images and image/signal processing}, \emph{meshes and grids}, \emph{charts/diagrams/plots}, \emph{queries and search}, \emph{comparison/comparative visualization}, \emph{interpolation}, \emph{motion}, \emph{programming}, and \emph{streamlines/pathlines/streaklines}.

\subsection{Analysis of Individual Keywords}
\label{sec:individualkeywords}
In this section, we analyze individual keywords in more detail across all three of our data sources. 

\subsubsection{Common Keywords}
The top three most frequently occurring keywords in the author keyword list were community-related keywords: \emph{information visualization}, \emph{visual analytics}, and \emph{visualization} (in this order). These are all keywords that describe a research (sub)community and it is interesting to see that the term scientific visualization does not occur. The terms \emph{information visualization} and \emph{visual analytics} were predominantly present on papers accepted to the respective conferences. We hypothesize that both IEEE VAST and IEEE InfoVis have been the smaller of the three conferences and in the process of becoming more established, authors may have chosen to tag their papers with these terms to strengthen associations to their respective communities. This hypothesis is also supported by the observation that \emph{information visualization} was the most frequent keyword for the period of 2004--2008, while \emph{visual analytics} was the most frequent keyword for period of 2009--2013 during which IEEE VAST became an established venue.

Next, we analyzed the top individual keywords per visualization conference to get an understanding of main topics and concerns in these subcommunities. We excluded the community-related keywords from this analysis. For author-assigned keywords, the top three for each IEEE VIS conference were:

\begin{description}[\compact\setlabelphantom{\emphbf{Vis/SciVis:}}]
\item[\emphbf{VAST:}] \emph{sensemaking} (13), \emph{visual knowledge discovery} (8), \emph{geovisualization} (7)
\item[\emphbf{InfoVis:}] \emph{interaction} (25), \emph{graph visualization} (19), \emph{evaluation} (17)
\item[\emphbf{Vis/SciVis:}] \emph{volume rendering} (72), \emph{flow visualization} (48), \emph{isosurfaces} (25)
\end{description}

Among these, no common keywords exist between the conferences. When expanding the analysis to the top 10, VAST and InfoVis share the keyword \emph{interaction} but no other overlaps emerged. This may indicate that the three conferences do focus on different areas of visualization---or use different terminology for related concepts. 

In order to understand whether terminology plays a role, we looked at the top three expert keywords. Here, the keyword \emph{interaction} emerges for both IEEE InfoVis and VAST but remains the only shared keyword: 

\begin{description}[\compact\setlabelphantom{\emphbf{Vis/SciVis:}}]
\item[\emphbf{VAST:}] \emph{analysis process} (53), \emph{interaction} (31), \emph{applications} (24)
\item[\emphbf{InfoVis:}] \emph{graphs} (73), \emph{interaction} (69), \emph{evaluation} (53)
\item[\emphbf{Vis/SciVis:}] \emph{volume rendering} (148), \emph{hardware and computation} (89), \emph{flow} (70)
\end{description}

When expanding the analysis to the top 10 keywords, \emph{application} emerges as a keyword common to all three conferences. InfoVis and VAST also share the keyword \emph{evaluation} (ranked 14 in Vis/SciVis) while VAST and Vis/SciVis share the keyword \emph{time} (ranked 12 in InfoVis). Among the top 100 keywords, 17 were used only on Vis/SciVis papers (ignoring a up to one occurrence at each of the other two conferences). The top 3 of these keywords were: \emph{volume rendering}, \emph{isosurfaces}, and \emph{geometric modeling}. We can conclude that all three sub-communities share some joint concerns, \eg on applications, evaluation, and temporal data but that they do have their respective foci. However, there appears to be more overlap between VAST and InfoVis than we observe between Vis/SciVis and the other two conferences according to the expert keywords.

The top three PCS keywords \emphbf{only} for the IEEE VisWeek/VIS were:
\begin{description}[\compact\setlabelphantom{\emphbf{Vis/SciVis:}}]
\item[\emphbf{VAST:}] \emph{visual knowledge discovery} (160), \emph{text and document data} (78), \emph{coordinated and multiple views} (74)
\item[\emphbf{InfoVis:}] \emph{graph/network data} (197), \emph{visual knowledge discovery} (99), \emph{user interfaces} (98)
\item[\emphbf{Vis/SciVis:}] \emph{volume rendering} (235), \emph{biomedical and medical visualization} (193), \emph{flow visualization} (154)
\end{description}

It can be seen that the top three keywords for the three datasets relate to each other in that \emph{graphs} and \emph{volume rendering} remain the top keywords for InfoVis and Vis/SciVis, respectively, and that \emph{visual knowledge discovery} was in the top 3 for VAST in the author keywords and also covered under \emph{analysis process} for the expert keywords. When looking at the top 100 PCS keywords, 6 emerged that were only used by submissions to Vis/SciVis (ignoring up to one occurrence at each of the other two conferences). These keywords all belong to the higher-level category of \emph{spatial data and techniques}: \emph{vector field data}, \emph{scalar field data}, \emph{extraction of surfaces}, \emph{tensor field data}, \emph{irregular and unstructured grids}, and \emph{volume modeling}. Conversely, there were two keywords never or only once used for submissions to Vis/SciVis: \emph{business and finance visualization} and \emph{tabular data}.

\subsubsection{Rising and Declining Keywords}
Next, we were interested in historic trends for individual keywords. We used Tableau to calculate linear trend lines for 
the 15 most frequently used keywords in the author, expert, and PCS datasets. \autoref{tab:expert-history} gives an overview of this data by dataset, ordered by slope from increasing to decreasing. Significant results are highlighted in bold (note that not all significant trends are shown due to the selection by frequency).

\textsc{Rising Keywords:}
Across all three IEEE VIS conferences, the use of the term \emph{interaction} is significantly increasing according to author-chosen keywords. If one only considers the expert keywords, the terms \emph{interaction}, \emph{evaluation}, \emph{multidimensional/multivariate}, and \emph{machine learning \& statistics} are on the rise. The analysis of the PCS keywords yielded 8 keywords that are significantly increasing: \emph{coordinate and multiple views, visual knowledge discovery, time-series data, data transformation and representation, high-dimensional data, interaction design, multidimensional data,} and \emph{graph/network data}. There is overlap between these increasingly popular topics. We attribute the strongest overall increasing trends to: \emph{interaction}, \emph{evaluation}, \emph{multi-dimensional}/\emph{high-dimensional}/\emph{multi-variate data}. 

\textsc{Declining Keywords:}
According to author-chosen keywords, both \emph{volume visualization} and \emph{flow visualization} have been significantly decreasing in frequency of occurrence over the past 10 years. The expert-coded keywords confirm this and add \emph{hardware and computation} as a declining topic. Interestingly, no significantly declining topics were found for the PCS taxonomy keywords (but \emph{GPUs} and \emph{multi-core architectures}---not shown in the table---trended that way, $p=0.07$). This may be due to the fact that we only had data for six years for PCS but 10 years for the author/expert keyword sets.

For the author/expert keywords, it is interesting to note that two very core and frequent keywords for the IEEE Vis/SciVis conference are in significant decline. This could perhaps be due to the fact the many fundamental research questions have been tackled and that researchers are now using more specific or other keywords.

\begin{sidewaystable}[p]
  \centering
  \footnotesize
	\tabulinesep=2pt
	\vspace{-\abovecaptionskip}\caption{Historical trends for 15 most frequently used keywords for each of the author, expert, and PCS datasets. Significant trends highlighted.}\vspace{.7\abovecaptionskip}
  \begin{tabu}{%
	@{}%
	X[l]@{\hspace{3pt}}r@{\hspace{3pt}}r@{\hspace{3pt}}r@{\hspace{3pt}}r%
	@{\hspace{3pt}}|@{\hspace{3pt}}X[1.15,l]@{\hspace{3pt}}r@{\hspace{3pt}}r@{\hspace{3pt}}r@{\hspace{3pt}}r%
	@{\hspace{3pt}}|@{\hspace{3pt}}X[1.75,l]@{\hspace{3pt}}r@{\hspace{3pt}}r@{\hspace{3pt}}r@{\hspace{3pt}}r%
	@{}%
	}
	\toprule
  \multicolumn{5}{@{}c@{\hspace{3pt}}|@{\hspace{3pt}}}{author (InfoVis, Vis/SciVis, VAST; 2004--2013)} & \multicolumn{5}{@{}c@{\hspace{3pt}}|@{\hspace{3pt}}}{expert (InfoVis, Vis/SciVis, VAST; 2004--2013)} & \multicolumn{5}{@{}c@{}}{PCS (EuroVis, PacificVis, InfoVis, Vis/SciVis, VAST; 2008--2013)} \\
	keyword & \# & slope & SE & p-val. & keyword & \# & slope & SE & p-val. & keyword & \# & slope & SE & p-val. \\
	\midrule
  \textbf{interaction} & 38 & 0.59 & 0.16 & \textbf{0.006} & \textbf{interaction} & 152 & 1.37 & 0.25 & \textbf{0.001} & \textbf{Coordinated and Multiple Views} & 323 & 10.71 & 3.24 & \textbf{0.030} \\
  visual analytics & 86 & 0.58 & 0.52 & 0.299 & \textbf{evaluation} & 105 & 1.32 & 0.42 & \textbf{0.014} & \textbf{Visual Knowledge Discovery} & 414 & 9.20 & 1.79 & \textbf{0.007} \\
  uncertainty visualization & 19 & 0.33 & 0.16 & 0.086 & \textbf{multidimensional/multivariate} & 83 & 0.87 & 0.34 & \textbf{0.034} & \textbf{Time Series Data} & 293 & 8.89 & 2.40 & \textbf{0.021} \\
  evaluation & 26 & 0.30 & 0.17 & 0.111 & visual analytics & 86 & 0.58 & 0.52 & 0.299 & \textbf{Data Transformation and Representation} & 252 & 7.77 & 2.46 & \textbf{0.034} \\
  user study & 22 & 0.10 & 0.12 & 0.416 & analysis process & 113 & 0.36 & 0.32 & 0.296 & \textbf{High-Dimensional Data} & 278 & 7.54 & 1.51 & \textbf{0.008} \\
  visualization & 81 & -0.01 & 0.30 & 0.984 & visualization techniques \& tools & 72 & 0.34 & 0.26 & 0.235 & \textbf{Interaction Design} & 245 & 6.66 & 2.28 & \textbf{0.043} \\
  parallel coordinates & 23 & -0.08 & 0.19 & 0.680 & graphs & 91 & 0.26 & 0.36 & 0.494 & Geographic/Geospatial Visualization & 251 & 5.69 & 2.23 & 0.063 \\
  clustering & 20 & -0.12 & 0.14 & 0.445 & data and data management & 64 & 0.24 & 0.31 & 0.450 & User Interfaces & 291 & 4.49 & 2.65 & 0.166 \\
  graph visualization & 21 & -0.13 & 0.12 & 0.321 & visualization & 81 & -0.01 & 0.30 & 0.984 & Biomedical and Medical Visualization & 379 & 3.80 & 2.06 & 0.139 \\
  focus+context & 18 & -0.17 & 0.12 & 0.195 & time & 85 & -0.13 & 0.35 & 0.723 & \textbf{Multidimensional Data} & 306 & 3.49 & 1.00 & \textbf{0.025} \\
  information visualization & 95 & -0.24 & 0.35 & 0.521 & information visualization & 95 & -0.24 & 0.35 & 0.521 & \textbf{Graph/Network Data} & 460 & 3.31 & 0.93 & \textbf{0.024} \\
  isosurfaces & 25 & -0.42 & 0.23 & 0.103 & applications & 103 & -0.27 & 0.30 & 0.392 & Volume Rendering & 446 & 2.40 & 3.88 & 0.569 \\
  \textbf{volume visualization} & 24 & -0.52 & 0.20 & \textbf{0.042} & \textbf{flow} & 74 & -0.70 & 0.24 & \textbf{0.019} & Visualization System and Toolkit Design & 274 & 2.34 & 3.06 & 0.487 \\
  volume rendering & 72 & -0.72 & 0.37 & 0.091 & \textbf{volume rendering} & 149 & -1.42 & 0.46 & \textbf{0.014} & Time-varying Data & 334 & 2.29 & 1.39 & 0.176 \\
  \textbf{flow visualization} & 48 & -0.82 & 0.26 & \textbf{0.016} & \textbf{hardware and computation} & 98 & -1.59 & 0.43 & \textbf{0.006} & Flow Visualization & 267 & 1.86 & 1.33 & 0.234 \\
	\bottomrule
  \end{tabu}
  \label{tab:expert-history}
\end{sidewaystable}

\subsection{Limitations}
\label{sec:limitations}
While our analysis has revealed a wealth of information, the study results have to be read in light of several analysis limitations.
One obvious limitation is, of course, that we only analyzed a subset of publications from the visualization domain. To determine this subset, however, we followed advice from Bradford's law \cite{Bradford:1985:SOI} for selecting our data sources. This law states that a small core of publications will account for as much as 90\% of the literature in terms of citations received---trying to achieve 100\% will add publications at an exponential rate. Thus, it is unreasonable to attempt to cover the whole literature for a field. Given the size and importance of IEEE VisWeek/VIS, we focused on a ten-year dataset from this conference (plus the additional data on the use of the PCS taxonomy). This analysis enabled us to get a rich overview of the field. Yet, compared to several past keyword analysis studies (\eg, \cite{ Callon:1986:MDS,Liu:2014:CMT,Liu:2012:CAD}), the visualization field is still young and the overall number of keywords was comparably low, in particular for the author-assigned keywords. The low number of overall keywords and the vast difference in number of papers accepted to the IEEE VisWeek/VIS conferences is also one of the main reasons we did not study the difference between IEEE VAST, InfoVis, and Vis/SciVis in depth but looked at the whole field together. By expanding the dataset back to 1995 (the first year of InfoVis) a comparative analysis may be more meaningful. Yet, another peculiarity of the field may further impact such a comparison: for a long period of time it was possible to send information visualization papers to both the InfoVis and Vis/SciVis conference. The Vis/SciVis conference even included information visualization sessions. Thus, the borders between the conferences were not always as clear as their current names may suggest.

For the PCS data we discussed a major limitation earlier that pertains to the different use of these keywords compared to author-assigned keywords. In addition, we also found that several older papers included ACM classifiers and it is possible that authors back then only selected author keywords in addition to these classifiers and did not provide duplicates. As we did not want to speculate on what may have happened, we  collected the author keywords as present on each paper.

The expert coding process that led to the expert keywords also, of course, includes an inherent subjective coding bias. While our team consisted of 5 experts with varying backgrounds in VAST, InfoVis, Vis/SciVis, and HCI, it is entirely possible that with other coders other clusters and cluster names may have emerged. Finally, this keyword analysis was conducted by experts in the field of visualization and not professional social scientists. Our own experience and views of our field have certainly tainted our interpretations of the data---as is common with any kind of qualitative analysis \cite{Norman:2005:HBO}.

\section{\href{http://www.keyvis.org/}{keyvis.org}: A Keyword Search Tool}
\label{sec:webpage}

To make our data accessible for others, we created a webpage  that makes author and expert keywords and related papers search- and browsable: \url{http://www.keyvis.org/}\,.
Visitors can search all 2629 unique author-assigned keywords, find out which keywords co-occurred how frequently, which manual expert clusters they belong to, and the actual research papers they appear on. 

Our main goal was to generate an easy-to-use, lightweight interface to our keyword data in order to: (a) support visualization researchers in making more informed decisions when picking keywords for their papers, and (b) give a new lens on identifying relevant related work. We have used the site ourselves for choosing keywords and finding related work that we were not aware of before. We hope that others will find it similarly useful.
In the long run, our goal is to maintain the website as a platform for visualization keyword access and analysis. 

\section{Discussion and How to Move Forward}
\label{sec:forward}

The analysis of the keyword data has revealed several major themes and declining and rising keywords in visualization. Based on the strategic diagrams, we have identified, in particular, motor themes of the field. 
While we have collected a large amount of author keyword data, invested heavily in clustering related author keywords, and compared this data to a standardized taxonomy, our analysis is only a first step in the direction of two larger research goals:

\textbf{Creating a common vocabulary:} 
In particular the analysis of raw author keywords has revealed that authors choose many variants of similar keywords based on: singulars and plurals (\eg, glyphs and glyph), abbreviated versions (\eg, DTI vs.\ diffusion tensor imaging; InfoVis vs.\ information visualization), spelling (multidimensional vs.\ multi-dimensional), or specificity (\eg, multi-dimensional vs.\ multi-dimensional data vs.\ multi-dimensional visualization). Such a diversity of terms may be a reflection of the diversity of influences on the visualization field---but is is not helpful, in particular when one wants to search for keywords or---like us---gain an overview of the research themes of a community. We hope that \href{http://www.keyvis.org/}{keyvis.org} will help paper authors find common terms and reflect on their keyword usage before submitting a camera-ready paper. In addition, one can think about the problem of creating a common vocabulary for visualization more broadly. By identifying key terms and providing clear definitions, sub-communities in visualization may be able to communicate more clearly about similar approaches and, this, in turn can help to also collaborate more effectively with people outside the community. Finally, a common vocabulary can also facilitate to more easily understand emerging and declining research trends within the field.

\textbf{Establish a comprehensive taxonomy of visualization research:}
One of the goals that we had initially set out to accomplish is not yet achieved. Perhaps the holy grail of a keyword analysis is to amount into a taxonomy of the analyzed field, in our case visualization research. This could serve two purposes. One the one hand, a taxonomy will help to better communicate ``what is visualization'' to other disciplines, \ie, researchers and practitioners not part of the VIS community. On the other hand, we are hoping to be able to facilitate the crucial step of matching reviewers with papers and grants such that the peer review process improves and new contributions is seen in the right context. 

Yet, how to exactly created and \textbf{maintain} a comprehensive taxonomy of visualization keywords is still an open question. As can easily be seen from an inspection of both the top IEEE and INSPEC terms in \autoref{tab:top-ten-2004-2013}, broader taxonomies are not very successful in capturing the diversity of the visualization field. The PCS taxonomy, in turn, has been developed by a team of dedicated experts with the goals to improve the reviewing system for visualization papers. Yet, what is the right process for maintaining and changing the taxonomy? The visualization field is evolving and, thus, a visualization taxonomy should be regularly updated. Should an analysis such as ours be used to find trends (and keywords representing these trends) that have not been captured? Would it be possible to automate our process without requiring experts to clean and code the data? Should certain keywords be split or merged as sub-areas increase or decrease in popularity? Does it make sense to keep keywords in the PCS taxonomy that are rarely used---or should the taxonomy provide a broad view of the field in order to capture its topic diversity? Even when one has answers to these questions, how would one choose the right level of granularity for keywords? For example, the keyword \emph{evaluation} emerged as a significantly increasing topic for our expert-clustered keywords but did not for the PCS keywords. The reason being that the topic is split into seven sub-topics in PCS, out of which none showed an increasing trend taken alone. Finally, should there be separate taxonomies: one for visualization as a whole (\eg, to conduct analyses on emerging and declining topics, topic coverage across subcommunities, etc.) and one for the submission process for academic research? The former taxonomy could be large and evolving while the latter would have to be reduced in size and remain stable across conferences for given time periods to remain manageable for papers chairs, editors, reviewers, and authors.

In summary, the 25\textsuperscript{th} anniversary of the IEEE Vis/SciVis conference and thus of the whole IEEE VisWeek/VIS conference series is a good moment to take a look back at the diverse and evolving history of our field. We managed to provide only a glimpse at the available data in this paper but hope that our discussion will spark others to discuss and consider the importance of keywords, common vocabularies, and taxonomies for the field as a whole.

\section*{Acknowledgments}We would like to thank Vassilis Kostakos for sharing details about the keyword analysis process for CHI keywords. We also thank Nenad Kircanski and Johanna Schlereth for their help with creating \href{http://www.keyvis.org/}{keyvis.org}. 

\bibliographystyle{abbrv-doi-doi}

\bibliography{RR-8580}

\end{document}